\documentstyle[aas2pp4]{article}
\begin{document}

\def\p{$\pm$}
\def\ab{$\sim$}
\def\kms{km s${}^{-1}$}
\def\beam{beam$^{-1}$}
\def\gusten{G\"{u}sten}
\def\radms{rad m$^{-2}$}
\def\yz{Yusef-Zadeh}
\def\anantha{Anantharamaiah}
\def\pel{G358.85+0.47}
\def\a{$\alpha$}
\def\l{$\lambda$}

\title{Discovery of a Non-Thermal Galactic Center Filament (G358.85+0.47)
Parallel to the Galactic Plane}
\author{Cornelia C. Lang\altaffilmark{1,2},
K.R. Anantharamaiah\altaffilmark{1,3}, N.E. Kassim\altaffilmark{4}, T.J.W. Lazio\altaffilmark{4}}\altaffiltext{1}{National Radio Astronomy Observatory, Box 0, Socorro,
NM 87801, email: clang@nrao.edu}\altaffiltext{2}{Division of Astronomy, 8371 Math Sciences Building, Box
951562, University of California at Los Angeles, LA, CA 90095-1562}
\altaffiltext{3}{Raman Research Institute, Bangalore 560 080, India} 
\altaffiltext{4}{Code 7213, Naval Research Laboratory, Washington
D.C. 20375-5351}

\begin{center}
To appear in {\it The Astrophysical Journal Letters}
\end{center}

\begin{abstract}
We report the discovery of a new  non-thermal filament, G358.85+0.47, the ``Pelican'', located
\ab225 pc in projection from SgrA, and oriented parallel to the Galactic plane. VLA continuum
observations at \l20 cm reveal that this 7\arcmin~(17.5 pc) structure
bends at its northern extension and is comprised of parallel
strands, most apparent at its ends. Observations at \l6 and \l3.6 cm reveal
that the Pelican is a synchrotron-emitting source and is strongly linearly
polarized over much of its extent. The spectral index of the filament
changes from \a$_{20/6}$=$-$0.8 to \a$_{6/3.6}$=$-$1.5. The rotation
measures exhibit a smooth gradient, with values ranging from $-$1000
\radms~to +500 \radms. The intrinsic magnetic field is well-aligned along the length of the filament. Based on these properties,
 we classify the Pelican as one of the non-thermal
filaments unique to the Galactic center. Since these filaments (most
of which are oriented perpendicular to the Galactic plane) are
 believed to trace the overall magnetic field in the inner Galaxy, the
Pelican is the first detection of a component of this field {\it
parallel} to the plane.  The Pelican may thus mark a transition
region of the magnetic field orientation in the inner kiloparsec of the Galaxy. 

\end{abstract}

\section{Introduction}
The inner hundred parsecs of the Galaxy contain a wealth of
 unusual radio structures. Prominent among these are unusual linear sources
 known as the non-thermal filaments (NTF's). The
 NTF's are typically long (10's pc) and narrow ($<$0.5 pc) structures
 with non-thermal spectra and strong linear polarization, indicating
 synchrotron emission. The
 intrinsic magnetic field orientation in each NTF is aligned with its long
axis.  Sub-structure, in the form of splitting filaments, is observed in
nearly every case. In addition, both ionized and molecular gas appear
 to be associated with almost
every NTF (Morris \& Serabyn 1996). The most famous of the NTF's, the Radio Arc, is a bundle of filaments
 which extends for \ab40 pc at $\ell$=0\fdg2, b=0\fdg0 (\yz~\& Morris
 1987).  Other examples include the Northern and Southern Threads
 (Morris \& \yz~1985, \anantha~et al. 1991, Lang, Morris \& Echevarria
 1999), SgrC (Lizst \& Spiker 1995), G359.54+0.18 (\yz~et al. 1997),
 and the Snake (Gray et al. 1995).   

Although it is clear that the NTF's are magnetic in nature, the origin
of the relativistic electrons and their mechanism for acceleration
remain unclear. Several models of NTF generation have been proposed (see Morris 1996).  A common feature of these models is
that the NTF's trace the magnetic field in the Galactic center, giving
insight to the overall field geometry in this region (Morris 1994). As
the NTF's are all oriented roughly perpendicular to the Galactic plane
(within \ab20\arcdeg), it appears that the magnetic field in this
region is polodial, as opposed to the azimuthal field which traces the
spiral arms in most galaxies (Beck et al. 1996). In addition,
Yusef-Zadeh \& Morris (1987) argue that the field strengths in the
NTF's must be of order \ab1 mG, to explain the rigid and ordered
linear extents of the NTF's in the presence of the extreme turbulence
found at the Galactic center.

Recently, a new wide-field image of the Galactic center at \l90 cm was
produced by Kassim et al. (1999) based on the VLA data of \anantha~et al. (1991). In a careful examination of this image,
a previously unidentified linear feature was discovered \ab1\fdg5 from
SgrA (or 225 pc in projection assuming D$_{GC}$=8.0 kpc; Reid
1993). Close reinspection of the 843 MHz MOST survey image (Gray 1994) confirms the presence of this feature in those data, although the coarse resolution (\ab1\arcmin) makes it difficult to derive any spectral index information.
This feature has now been labeled \pel, based on its Galactic
coordinates.  \pel~has a linear structure similar to the NTF's, but it
is oriented parallel to the Galactic plane. In contrast, all other NTF's
are oriented perpendicular to the Galactic plane.  This paper reports
\l=20, 6, and 3.6 cm observations of \pel, in both total intensity and linear
polarization.  Initial results from \l20 cm VLA observations
were reported in \anantha~et al. (1999). Here, we demonstrate that
\pel~is properly classified as an NTF, and discuss the
implications of an NTF oriented parallel to the Galactic plane. 
  
\section{Observations \& Results}
All observations of \pel~were made with the VLA using a phase center at (\a,
$\delta$)$_{B1950}$=17 37 48.7, $-$29 38 17.0.  The observing
frequencies were 1.365 \& 1.435 GHz in the \l20 cm band, 4.585 \&
4.885 GHz in the \l6 cm band, and 8.085 \& 8.465 GHz in the \l3.6 cm
band.  The VLA was used in the B-array (\l20 cm), C-array (\l6 cm),
and DnC-array (\l3.6 cm), resulting in an approximately equal beamsize
(8\farcs5 $\times$ 4\arcsec) for all three bands.  At each frequency,
observations were made in dual polarization mode with a bandwidth of 50
MHz. Standard AIPS procedures were used for calibration, editing, and
imaging of all data. 3C 286 was used in all instances for flux
calibration, and NRAO 530 (1733$-$130) and 1748$-$253 were used for
phase and polarization calibration.

\subsection{Morphology of \pel, ``the Pelican''}

Figure 1 shows the \l20 cm continuum image of \pel, which appears as
a linear feature extending for 7\arcmin~(17.5 pc) parallel to the
Galactic plane.  At a projected distance of 225 pc from SgrA, this
filament is the most distant NTF-like feature from the Galactic
center. With the exception of its orientation, the morphological
characteristics of \pel~are similar to those of other NTF's.  \pel~is comprised
of multiple, parallel filaments, most apparent at its ends, and has
the appearance of a ``Pelican'' figure; hence the name. Three strands
can be distinguished at the SW end, and at least two strands are
distinguishable at the NE extent. One of the strands starts
\ab3\arcmin~NE of the center of the Pelican and appears to continue
through the center, forming the middle SW strand.  Similar to other
NTF's, the peak emission occurs at the center  of the filament. 
The NE extension of the Pelican has a different orientation by \ab45\arcdeg~from the orientation of the rest of the
filament; bendings reminiscent to this have only been observed in one other NTF, the Snake (Gray et
al. 1995). Observations of the Pelican at \l6 and \l3.6 cm reveal that
it has essentially the same structure as at \l20 cm.     

\vbox{%
\begin{center}
\leavevmode
\hbox{%
\epsfxsize=7.5cm
\epsffile{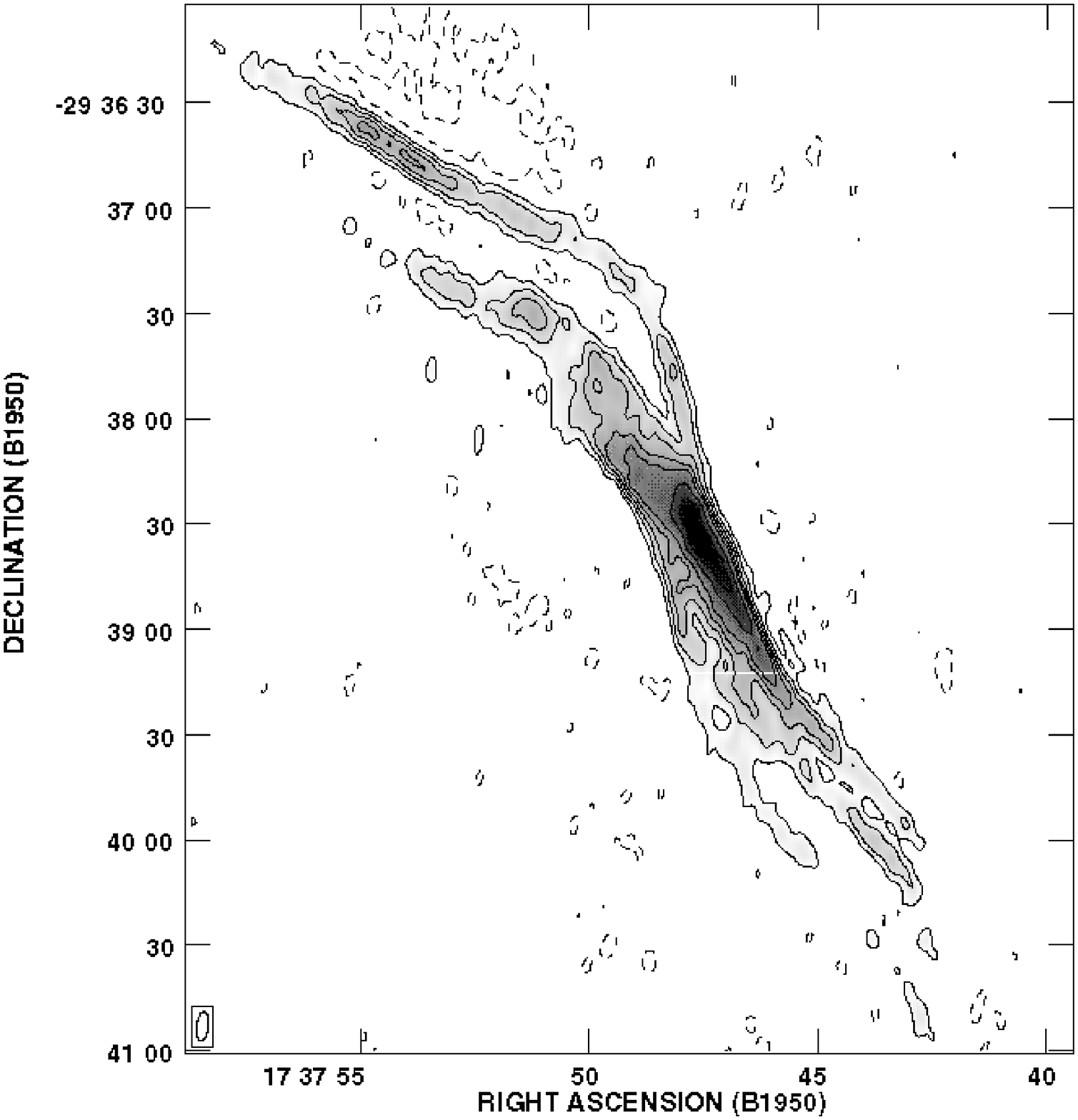}}
\begin{small}
\figcaption{\small 
\l20 cm continuum image of the Pelican shown in both contours
and negative greyscale. The resolution of this image is 8\farcs43
$\times$ 3\farcs39, PA=$-$7\arcdeg, and it has been corrected for primary
beam attenuation. The rms noise in this image is 55 $\mu$Jy
\beam. Contour levels represent -0.2, 0.2, 0.4, 0.6, 0.8, 1, 1.4, 1.8,
2.0, 2.2 mJy \beam.}
\end{small}
\end{center}}

\subsection{Spectral Index}

Based on the integrated fluxes at \l90 cm (Kassim et al. 1999) and
\l20 cm, the spectral index of the Pelican is \a=$-$0.6
(where S$_{\nu}$ $\propto$ $\nu^{\alpha}$), consistent with
synchrotron emission.  At \l=20, 6, and 3.6 cm, where the structure of
the Pelican is almost identical, pair-wise spectral indices were
determined using cross-cuts of intensity at several positions along
the length of the filament. The spectrum of the Pelican becomes
steeper with frequency: \a$_{20/6}$=$-$0.8\p0.2, and
\a$_{6/3.6}$=$-$1.5\p0.3, but the spectral index at each frequency
pair is constant as a function of position along the filament. Since
the beam diameter is essentially equal at each wavelength, the
observed steepening in the spectral index is unlikely to be
instrumental. An apparent flattening of the spectral index due to
free-free absorption at the longer wavelengths can be ruled out by 
the estimated free-free optical depth at \l90 cm.  If the actual
spectral index is that measured between \l6 and \l3.6 cm
(\a=$-$1.5), then in order to produce the observed \l20 cm flux
density, an optical depth of $\tau_{20cm}$=0.9 is required.  This
implies $\tau_{90cm}$=18, which is impossible given that the Pelican
was discovered at this wavelength. An alternate explanation is that
the observed steepening is due to an intrinsic break in the
spectrum. The dramatic steepening at shorter wavelengths may correspond to a
very sharp cutoff in the electron energy distribution. A similar
steepening of the spectral index has been observed in the Northern
Thread, where \a$_{20/6}$=$-$0.5, and \a$_{6/2}$=$-$2.0 (Lang et
al. 1999).  
 
\subsection{Polarization}
Observations of the Pelican at \l6 and \l3.6 cm were the most sensitive to
polarization.  At \l20 cm, for rotation measures $>$ 200 \radms, Faraday rotation across the 50 MHz bandwidth ($>$ 90\arcdeg)
will depolarize most of the polarized emission. A very low level of
polarization is detected at \l20 cm where the RM are $<$ 200 rad \radms, but
since it covers such a small portion of the Pelican and has low S/N,
we rely on the \l6 and \l3.6 cm data for polarization results. Figure
2 shows the distribution of polarized intensity at \l6 cm.  
The polarized emission is concentrated along a
central ridge of the Pelican, where several peaks of polarized emission
are obvious, and extends over more than half the length of the
filament. Polarized emission also arises from the NE extension of
the Pelican. Cross cuts of the polarized and total intensity at
several positions along the filament length were compared.  At \l6
cm, fractional polarizations are typically 50$-$60\%, whereas at
\l3.6 cm, the fractional polarization is 60$-$75\%.  Faraday rotation may
cause the \l6 cm fractional polarization to be slightly lower
than at \l3.6 cm, since the magnitude of Faraday rotation is
proportional to \l$^2$. The fractional polarization in the Pelican is
higher and more coherent across the filament than in other NTF's (Gray et al. 1995, \yz~et al. 1997, Lang et al. 1999).  As the
fractional polarizations are near the theoretical limit for
synchrotron emission (75\%), there does not appear to be any
significant depolarization toward the Pelican. At \l6 cm and \l3.6
cm, bandwidth depolarization is insignificant for $\Delta$$\nu$=50
MHz.  Beam depolarization can also not be a significant effect
since we find that the intrinsic magnetic field is highly ordered (see
below).  

\vbox{%
\begin{center}
\leavevmode
\hbox{%
\epsfxsize=7.5cm
\epsffile{lang.f2.cps}}
\begin{small}
\figcaption{\small 
Total intensity and polarized emission at \l6 cm arising
from the Pelican, shown in RA and DEC units of arcminutes offset from
the phase center (\a, $\delta$$_{B1950}$=17 37 48.7, $-$29 38 17.0). The contours show the total intensity with a
resolution of 8\farcs36 $\times$ 3\farcs54, PA=$-$1\arcdeg, and an rms
noise level of 20 $\mu$Jy \beam. The contour levels represent 0.1, 0.2, 0.4, 0.8 mJy \beam. The polarized emission
 shown
in false-color scale with a resolution of 9\arcsec $\times$5\arcsec,
in the range of 150$-$900 $\mu$Jy \beam. }
\end{small}
\end{center}}

\subsection{Rotation Measure \& Intrinsic Magnetic Field Orientation}
Using the polarization angles at the four observed frequencies (4.585,
4.885, 8.085, \& 8.465 GHz), we can solve for the Faraday Rotation
toward the Pelican. Figure 3 shows the rotation measure (RM)
distribution, with RM values in the range of $-$1000 to +500
\radms.  Errors were estimated using fits of the rotation angle to
\l$^2$, and are in the range of \p10 \radms. There is a strong gradient of RM across the Pelican; at the NE
extent, RM=500 \radms, at the center, RM=0 \radms, and at the SW end,
RM=$-$500 \radms.
  The RM
values toward the Pelican are less than those toward the other NTF's,
where typically rotation measures are 2000$-$5000 \radms~(Gray et
al. 1995, \yz~et al. 1997, Lang et al. 1999). The sign
reversal of the RM toward the center of the Pelican has also not been
observed toward any of the other NTF's.  Both the high fractional
polarization and the \l$^2$-dependence of the rotation angle
indicate that internal Faraday rotation is not occuring in the
Pelican. Therefore, the sign reversal in the RM implies that the
magnetic field along the line of sight to the Pelican must
undergo a sign reversal in the intervening medium.
\vbox{%
\begin{center}
\leavevmode
\hbox{%
\epsfxsize=7.5cm
\epsffile{lang.f3.cps}}
\begin{small}
\figcaption{\small 
 Distribution of rotation measure along the Pelican shown in
false-color scale with units of rad m$^{-2}$, superposed with the 0.2
mJy \beam~contour of total \l20 cm intensity from Figure 1. The RA and
DEC axes are the same as in Figure 2.}
\end{small}
\end{center}}

Figure 4 shows the orientation of the intrinsic magnetic field in the
Pelican, after correction for Faraday rotation.  The
magnetic field is remarkably well aligned along the entire extent of
the filament, showing that the structure is indeed dominated by the
magnetic field. Therefore, the highest equipartion magnetic field
value obtained for
the Pelican in its narrow NE extension, \ab70 $\mu$G, is a lower limit. At the center of the Pelican, the orientation of the
field is parallel to the filament and to the Galactic plane;  in the
NE extension, where the filament bends by \ab45\arcdeg, the field orientation also
changes by \ab45\arcdeg. The magnetic field orientation in both
strands of the NE region of the Pelican bend as the
filament does, thus demonstrating that the magnetic field also dominates the
sub-structure of the filament.  Figure 4 therefore
illustrates the first detection of a large-scale magnetic field in the
Galactic center region that is oriented differently from the
known magnetic structures. 
\vbox{%
\begin{center}
\leavevmode
\hbox{%
\epsfxsize=7.5cm
\epsffile{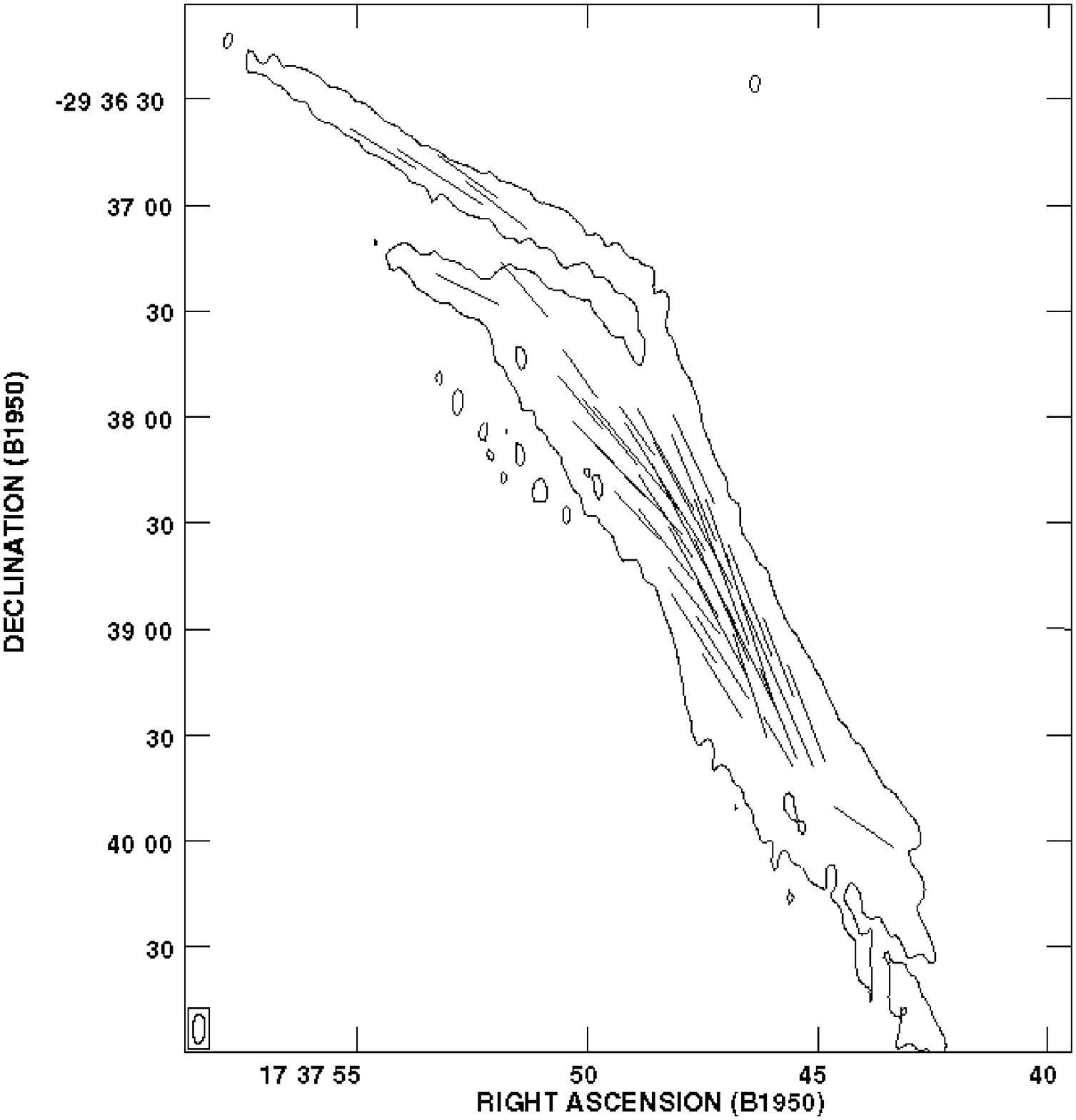}}
\begin{small}
\figcaption{\small 
 Vectors representing the intrinsic orientation of the
magnetic field in the Pelican are shown superposed on the 0.2 mJy
\beam~contour of \l6 cm continuum emission from Figure 2.}
\end{small}
\end{center}}

\section{Discussion}
There is no doubt that the Pelican should be
classified as a member of the unusual class of Galactic center
NTF's, despite its orientation parallel to the Galactic plane. Its morphology, spectral index, fractional polarization, and
alignment of the magnetic field along its length are all very
consistent with the known properties of the NTF's. The multiple
parallel strands in the Pelican are reminscent of the filaments in the
Radio Arc (\yz~\& Morris 1987), where more than a dozen
strands run along the length of the Radio Arc and appear to bifurcate
into multiple, parallel strands.  In the case of the Pelican, the filamentation manifests itself
on a smaller scale and there are fewer strands. The concentration of
brightness toward the center of the Pelican, and the wispy nature of
the strands of filamentation toward its end can be compared to
structure in SgrC, G359.54+0.18 and G359.8+0.2, which also show bright
centers and diffuse subfilamentation at their ends (\anantha~et
al. 1991, Lizst \& Spiker 1995, \yz~et al. 1997, Lang et al. 1999). 

Many of the well-studied NTF systems have associated ionized and
molecular gas, and there seems to be evidence for interaction between
them (Serabyn \& \gusten~1991, Serabyn \& Morris 1994, Uchida et
al. 1996, Staghun et al. 1998).  Such associations have in fact been
used to construct models for the generation of NTF's through the
interaction of the strong and ordered large scale magnetic field
(traced by NTF's) with the magnetic field that is tied to the partially
ionized molecular gas.  Where these field systems intersect, magnetic
field reconnection has been invoked to accelerate the elctrons along
the field lines, thereby illuminating the NTF's (Serabyn \& Morris
1994). The Pelican does not appear to be associated with any ionized
region, and any association with molecular material remains to be
investigated.  The relative isolation of the Pelican could thus prove
to be an interesting counterexample to the known NTF's, and may
constrain existing models for NTF generation. 

As the NTF's are the main probe of the large scale magnetic field
at the Galactic center, their structure and orientation are crucial for
understanding the magnetic field geometry in this
region. Morris (1994) points out that the gentle curvature of several
of the NTF's is suggestive of a dipolar field which diverges above and
below the plane.  The 45\arcdeg~bend in the northern filamentary
strands of the Pelican may establish a link between the magnetic field
orientation it traces and the perpendicaular component which the other
NTF's define. Indeed, there may be a transition region occuring at
\ab200 pc from the Galactic center, inside of which the magnetic
field appears to be vertical and highly ordered.
The Pelican demonstrates that the magnetic field in the inner regions of the Galaxy has a
more complicated structure than the apparently simple vertical dipolar
field, suggested by the known NTF's. 

\section{Summary}

We report the discovery of a new filamentary feature, \pel, located
\ab225 pc in projection from the Galactic center.  Based on its
appearance, we refer to it as ``the Pelican''.  Multifrequency VLA
observations show that the Pelican can be properly classified as a
Galactic center NTF.  The Pelican has a non-thermal spectrum and is strongly linearly
polarized, similiar to other NTF's. The unique feature of the Pelican
is its orientation, parallel to the Galactic plane, in contrast to
the rest of the NTF's which are oriented perpendicular to the plane.
After correction for Faraday rotation, the intrinsic magnetic field of
the Pelican is remarkably well-aligned
along its length. The N strand of the Pelican bends by
\ab45\arcdeg~from the orientation of the rest of the filament, and the
orientation of the intrinsic magnetic field follows this bend. Since
the NTF's are believed to trace the overall magnetic field configuration in the
inner few hundred parsecs of the Galaxy, the detection of the Pelican is
the first component of the large scale magnetic field found in the
inner Galaxy to be parallel to the Galactic plane.

\section{Acknowledgements}
The Very Large Array (VLA) is a facility of the National Science
Foundation, operated under a cooperative agreement with the Associated
Universities, Inc. Basic research in radio astronomy at the Naval
Research Lab is supported by the office of Naval Research.

\end{document}